\documentclass[a4paper,11pt,amssymb,showkeys,
showpacs,eqsecnum,amsfonts,epsf]{article}

\usepackage{geometry}
\usepackage[dvips]{graphicx}
\usepackage{bm,latexsym,amsmath,amssymb,amsfonts}
\usepackage[usenames,dvipsnames]{color}
\usepackage[colorlinks=true,linkcolor=blue]{hyperref}
\usepackage{ulem}
\usepackage{epsfig}
\usepackage{kotex}

\definecolor{mypink1}{rgb}{0.858, 0.188, 0.478}
\definecolor{mypink2}{RGB}{219, 48, 122}
\definecolor{mypink3}{cmyk}{0, 0.7808, 0.4429, 0.1412}
\definecolor{mygray}{gray}{0.6}
\definecolor{pptbg}{rgb}{0.961,0.945,0.863}

\newcommand{\be}[1]{\begin{equation} \label{#1}}
\newcommand{\ee}{\end{equation}}
\newcommand{\bex}{\begin{equation*}}
\newcommand{\eex}{\end{equation*}}
\newcommand{\bea}{\begin{eqnarray}}
\newcommand{\eea}{\end{eqnarray}}
\newcommand{\ba}{\begin{array}}
\newcommand{\ea}{\end{array}}
\newcommand{\nn}{\nonumber}

\newcommand{\bel}{\begin{align}}
\newcommand{\eel}{\end{align}}

\newcommand{\mubb}{\mu}

\title{Collision of two spinning billiard balls and the role of table}

\author{Hyeong-Chan Kim \\
{\small School of Liberal Arts and Sciences, Korea National University of Transportation, Chungju 380-702, Korea }}

\begin{document}

\maketitle

\abstract{
We study the collision dynamics of a spinning cue ball approaching a static object ball with equal mass on a plane, common in billiards. 
While typical collisions in billiards are nearly perfectly elastic, with a restitution coefficient close to 1 and low friction, we explore three deviations from ideal elastic collisions:
The non-elastic nature, the friction effects between the balls during collision, the friction between the ball and the table.
We describe the detailed collision outcomes, emphasizing the importance of considering frictions. 
We reveal that friction, both between the balls and with the table, significantly influences the post-collision motions, deviating from the expectations of a purely elastic collision.
The insights gained contribute to a better understanding of ball dynamics, impacting strategies and gameplay in billiards.
}

\section{Introduction}
Billiard players claim that when a cue ball having topspin collides head on with an object ball, the cue ball momentarily moves backward before advancing due to friction with the floor. 
Individuals familiar with the perfectly elastic collision of two objects with equal mass find it difficult to intuitively accept this claim. 
However, upon examining super-slow-motion videos, such as those on Dr. Dave's YouTube channel~\cite{Dave}, it is confirmed that, immediately after the collision, the cue ball briefly lifts off the ground, moves slightly backward, and then advances again due to rotation. 
In this way, collisions between two billiard balls exhibit phenomena that deviate from intuitive expectations based on perfectly elastic collisions.

To understand these phenomena, a comprehend understanding of collision dynamics is necessary. 
Theoretical studies on billiards have seen limited development since the early 1830s when Coriolis authored a book on the mathematical theory of spin friction and collision in the game of billiards, translated into English by Nadler~\cite{Coriolis}. 
Nearly a century later, Moore attempted the second physics analysis~\cite{Moore}. 
Subsequent research has covered various aspects, including collisions between billiard balls~\cite{Walllace88, Crown}, collisions with cushions~\cite{Han2005}, high-speed camera analyses~\cite{Mathavan1, Mathavan2010}, and model studies for developing robotic systems for billiards~\cite{Nierhoff}. 
When using the cue to strike a billiard ball, the ball moves in a direction almost identical to the cue's striking direction. 
However, due to friction, the ball deviates slightly from the cue's direction, known as the squirt phenomenon~\cite{Alciatore TP21, colostate, Shepard, Cross2008, Cross2002, Stronge2018, hk1}. 
Although the angle is small, this deviation is a crucial factor that makes it challenging to hit a distant ball with the desired thickness in actual gameplay.

The motion of a billiard ball is described by the translational motion of the center and the rotational motion around the center. 
When billiard players use the cue to strike the ball, the ultimate goal is to control the translational and rotational motion appropriately. 
Additionally, after the cue ball collides with the first object ball, the aim is to guide both balls' trajectories according to the desired outcome. 
Precisely placing the cue ball at the intended position on the object ball requires extensive training. 
However, in practical games like 3-cushion or 4-cushion billiards, using the cue to send the cue ball to the object ball (assuming not aiming for a perfectly accurate thickness) is relatively straightforward, excluding the squirt/curve phenomenon. 
Therefore, during the game, the most critical element is understanding how the moving ball (cue ball) will behave after colliding with the stationary ball (object ball). 
This study aims to address this aspect.

Firstly, two approximations that closely hold during the collision of two billiard balls are presented to assist in calculations. 
First, the friction between the balls during collision is considered negligible. 
The friction coefficient between billiard balls, denoted as $\mu$, varies depending on the state and type of the balls but generally exists in the range of 
\be{mu1}
0.03 \leq \mubb \leq 0.08 . 
\ee
Assuming no friction, angular momentum is not transferred between the colliding balls. 
Secondly, billiard balls undergo a perfectly elastic collision. 
The coefficient of restitution during the collision is approximately
\be{e*}
0.92 \leq e_{*} \leq 0.98.
\ee
\begin{figure}[tbh]
\begin{center}
\begin{tabular}{c}
\includegraphics[width=.9\linewidth]{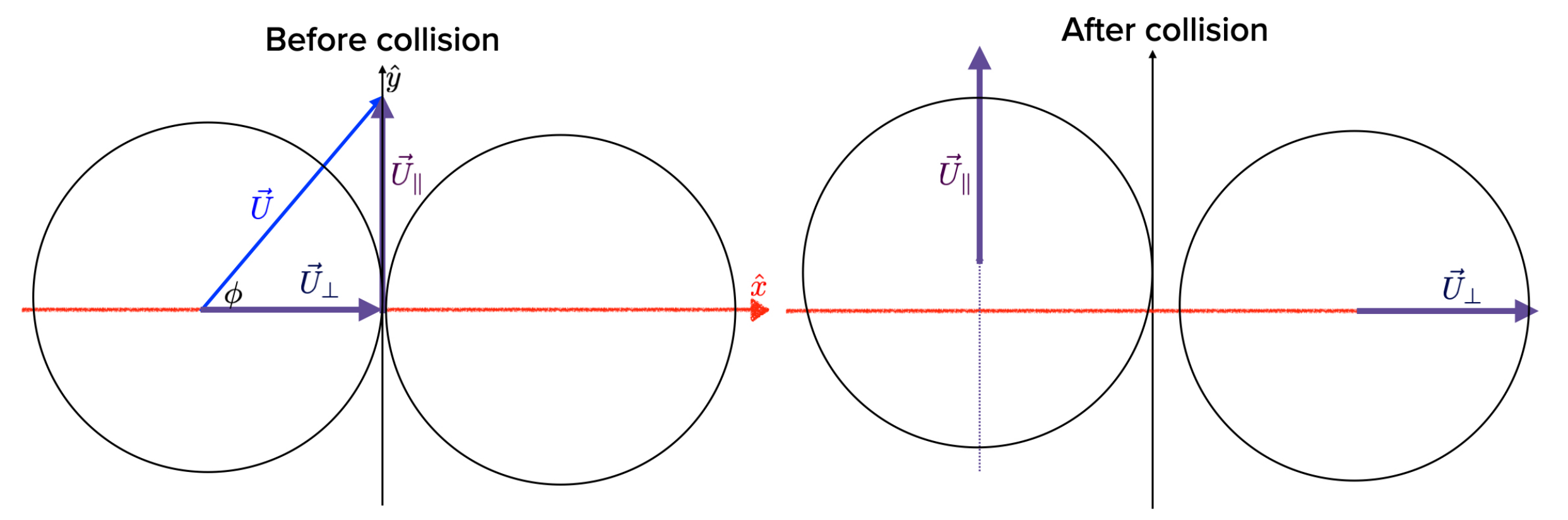} 
\end{tabular}
%
%
\end{center}
\caption{\footnotesize 
The perfectly elastic collision between two balls. The velocities of the two balls perpendicular to the collision plane are exchanged. In this illustration, the collision plane is represented by the vertical line. After the collision, the two balls move perpendicular to each other.}
\label{fig:vt}
\end{figure}
A collision with a restitution coefficient of 1 is considered a perfectly elastic collision. 
Due to the coefficient of restitution being very close to 1, the collision between the balls closely resembles a perfectly elastic collision. 
For example, colliding a non-rotating cue ball head-on with a stationary object ball results in the cue ball to stop, and the object ball to move almost at the same speed as the cue ball before the collision. 
Furthermore, colliding a non-rotating cue ball with a non-zero collision parameter with a stationary object ball shows that the separation angle between them immediately after the collision is very close to $90^{\rm o}$, as depicted in Figure~\ref{fig:vt}. 
Although the actual separation angle is slightly less than $90^{\rm o}$ due to the collision not being perfectly elastic, it is typically around $85^{\rm o}\sim 89^{\rm o}$ and does not significantly impact real games. 
It is clear that the collision between balls is never perfectly elastic as sound and heat are generated during the collision. 
However, in practical gameplay, starting with the approximation of frictionless perfectly elastic collisions and making corrections as needed is often sufficient.

If gravity is disregarded, three factors are involved in the adjustment: friction between the colliding balls (proportional to the friction coefficient $\mu$), deviation from a perfectly elastic collision ($e_*-1$), and friction between the ball and the table. 
Recent studies by Peskin~\cite{Peskin} considered the collision of rotating billiard balls, taking into account the friction between the balls. 
However, no solutions considering the effects of inelastic collisions and friction with the floor currently exist.

The billiard table serves two roles during the collision of balls. 
Firstly, it provides normal forces to the balls during the collision that prevent them from falling. 
Secondly, it imparts frictional force to the balls, proportional to the vertical normal force. 
Frequently, it was assumed that significant changes in the vertical normal force do not occur before the collision ends to ignore the effect of this frictional force. 
If this assumption holds, the collision between two balls can be much simplified. 
On the contrary, if this assumption is incorrect, friction with the floor will affect the post-collision motion of the balls. 
In reality, billiard balls roll on a smooth cloth laid over a hard surface. 
Therefore, the object providing vertical support force to the balls is the hard surface, and the cloth plays the role of providing friction with the balls.

The collision time between two colliding billiard balls is approximately $250\, \mu \text{s} < \Delta t < 300 \,\mu\text{s}$. 
Let the cue ball's velocity before the collision be $U_i$ and its mass be $M$.
Then, the impact force of approximately $M U_i/\Delta t$ acts between the balls during the collision. 
This force, delivering an impact roughly equivalent to the magnitude of momentum during the collision time, is approximately $400$ times greater than the force of gravity acting on the balls. 
Using the friction coefficient between the balls~\eqref{mu1} ($\mu \sim 0.05$), the magnitude of the frictional force  ($f\sim \mubb F_\perp$) between the balls due to rotation is approximately $20$ times greater than the gravity. 
Only when the cue ball collides with the object ball at a very low speed ($\sim 5\,\text{cm/s}$), the frictional force's magnitude becomes comparable to gravity. 
When friction acts in the direction of lifting the cue ball upward, the force pressing the cue ball against the table disappears.
Consequently, the vertical normal force exerted by the table on the cue ball would become zero. 
The magnitude of the impulse imparted to the cue ball by this frictional force is approximately \( f \Delta t \sim \mubb F_\perp \Delta t = \mubb M U_i \), and the resulting vertical height gained by the ball, regardless of collision time, is approximately given by:
\[ h = \frac{\mubb^2 U_i^2}{2g} \]
For a collision with an initial speed of \( U_i = 10 \, \text{m/s} \) and a friction coefficient of \(\mubb \approx 0.05\), this yields \( h \approx 1.3 \, \text{cm}\), which is an observable value. However, for slower cue ball speeds, such as \( U_i \approx 1 \, \text{m/s} \), the height \( h \) becomes \( \sim 10^{-2} \, \text{cm} \), making it practically imperceptible. 
In such a situation, if the assumption mentioned in the previous paragraph is correct, the object ball would move in the downward direction with the same speed. 
Consequently, the object ball would collide with and rebound off the surface of the table instantly.
In typical situations, when the cue is used to strike the cue ball horizontally, causing the two balls to collide with equal heights, one will observe that one of the balls (either the cue ball or the object ball) tends to rise, while the other moves horizontally\footnote{Note that in certain cases during a game, the cue may be tilted to intentionally create a height difference between the two balls during collision, resulting in a more significant vertical impulse.}, depending on the direction of rotation, which is contrary to the previous expectation.

These observations suggest that the assumption may not be appropriate depending on situation that significant changes in the vertical normal force do not occur before the collision ends to ignore the effect of the frictional force with the table so that the role of the table is negligible in the collision. 
On the other hand, if the frictional force lifts the cue ball, the reactional force tends to push the object ball downward. The magnitude of this frictional force is significantly greater than gravity, making it impossible to ignore. 
Then, the normal force exerted by the floor on the object ball cannot be neglected either.
Additionally, the frictional force between the object ball and the table would also large, particularly due to the static friction present when the object ball is at rest before the collision.
 The static friction between the table and the ball, which brings the ball to a stop, is determined by the material of the table surface. 
 Since the collision occurs in a very short time, the fiber material cannot respond during the collision. 
 Therefore, the table and the ball experience the maximum static friction during the collision time. 
 This static friction is significantly larger than other frictional forces, making it a crucial factor that causes significant deviation from the results expected in a perfectly elastic collision.

\vspace{.2cm}
The equations of motion for the collision between two balls with equal masses will be formulated in Section II. Section III will demonstrate how these equations are utilized during the collision process. Section IV will analyze the collision results, and finally, Section V will summarize the research findings in this paper.

\section{Equations of motion for the two ball collisions on a plane} \label{sec:str}
\begin{figure}[tbh]
\begin{center}
\begin{tabular}{c}
\includegraphics[width=.9\linewidth]{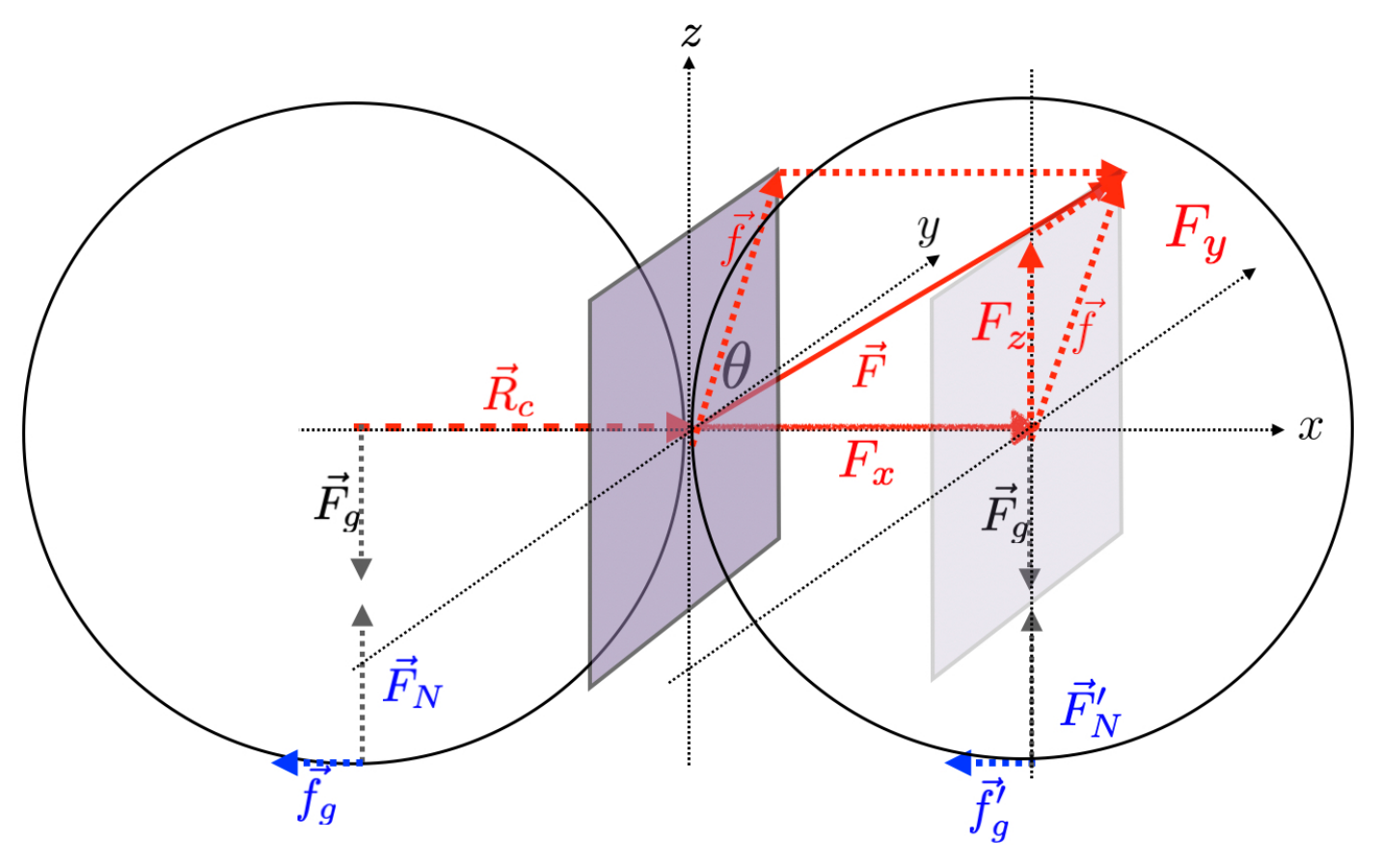} 
\end{tabular}
\end{center}
\caption{\footnotesize Various forces acting during the collision of the cue ball and the object ball. 
The cue/object ball is placed on the left/right.}
\label{fig:vt2}
\end{figure}

Let the mass of the cue ball and the object ball be denoted as $M$, and their radii as $R$.
Assume both balls move on a flat table, which is parallel to the $(x,y)$ plane.
Consider colliding the cue ball, which is in motion with a center velocity $\vec{U}_i = (U_{ix}, U_{iy}, 0)$ and angular velocity $\vec{\omega}_i$ before the collision, with the static object ball.
Assume that both balls are of the same size, the table is horizontal, and the balls do not rise before the collision; hence, the collision plane is perpendicular to the table, as shown in Figure~\ref{fig:vt2}.
Define the direction from the center of the cue ball through the collision point to the center of the object ball as the $x$-direction and the intersection of the collision plane (the $(y,z)$-plane in Figure~\ref{fig:vt2}) and the table as the $y$-direction.
Here, let the center velocity $\vec{U}_i$ of the cue ball before the collision be inclined by an angle $\phi$ with respect to the $x$-axis, as shown in Figure~\ref{fig:vt}.
Finally, choose the $z$-axis as the vertical upward direction perpendicular to the table.
This arrangement is illustrated in Figure~\ref{fig:vt2}.
In the following, we denote the center velocities of the cue ball and the object ball as $\vec{U}$ and $\vec{V}$, respectively, and their angular velocities as $\vec{\omega}$ and $\vec{\Omega}$.

As illustrated in Figure~\ref{fig:vt2}, the force $\vec{F}$ exerted by the cue ball on the object ball during the collision, consists of the sum of the force $\vec{F}_x = (F_\perp, 0, 0)$ in the $x$-direction perpendicular and the friction force $\vec{f}$ acting parallel to the collision plane.
The friction force acts in the opposite direction to the relative velocity of the collision point and its magnitude is generally given by the product of the friction coefficient $\mu$ and the normal force $F_\perp$, with its direction determined by the angle $\theta$ with the horizontal direction (choose the positive $y$-direction so that $-\pi/2 \leq \theta \leq \pi/2 $):
\be{f}
\vec{f} = (0, f\cos\theta, f\sin\theta), \qquad f = \mu F_\perp.
\ee
Thus, if there is topspin or draw on the cue ball ($\omega_y \lessgtr 0$), the angle $\theta$ satisfies $\theta \gtrless 0$.

Listing the forces acting on the cue ball, $-\vec{F} = (-F_\perp, 0, 0)-\vec{f}$ is the reaction to the action of the cue ball on the object ball (due to friction and normal force).
$\vec{f}_t$ is the friction on the cue ball exerted by the table, and $\vec{F}_N$ is the normal force supporting the cue ball on the table, respectively.
Finally, $\vec{F}_g$ represents the gravitational force acting on the cue ball.
As discussed earlier, the normal force $\vec{F}_N$ supporting the cue ball on the table must counterbalance the sum of the vertical forces due to gravity and friction with the object ball.
Therefore,
\be{FN}
\vec{F}_N  \equiv F_N \hat z ; \qquad
F_N=(M g + f \sin\theta) \Theta(\theta+\theta_c) ,
\ee
where $\theta_c \equiv \arcsin(Mg/\mu F_\perp)$ and $\Theta(x)$ is the Heaviside step function.
If the frictional force in the vertical direction is greater than or equal to the gravitational force ($ M g + \mu F_\perp \sin\theta< 0 $), the normal force becomes zero, and the cue ball moves upward due to the collision.
The friction force $\vec{f}_t$ on the table is given by the product of the normal force magnitude $F_N$ and the coefficient of friction between the ball and the table.
If the cue ball was rolling before the collision, the friction coefficient is very small ($\mu_{\rm rolling}\sim 10^{-2}$), and the friction force can be neglected during the rolling motion.
Assuming a smooth table surface compared to the cue ball, the friction force does not significantly affect the ball's motion before the collision ends.
If the collision occurs in a sliding state before the collision, the sliding friction coefficient $\mu_{\rm bt} \sim 0.1$ comes into play, and in this case, the frictional force is relatively large compared to the rolling case.
The magnitude of the friction force with the table is given by
\be{ft}
f_t = \mu_{\rm bt} F_N \sim \mu_{\rm bt} \mu F_\perp \Theta(\theta+\theta_c).
\ee
Depending on the situation, if the vertical force becomes very large, the friction with the table cannot be ignored. However, it still acts as a second-order approximation since its magnitude is proportional to the product of the two friction coefficients. Therefore, in this paper, we assume that the friction between the cue ball and the table can be ignored compared to $F_\perp$.

Similarly, for the object ball, the forces include the force exerted by the cue ball $\vec{F}$, the friction force from the table $\vec{f}'_t$, and the normal force supporting the object ball $\vec{F}'_N$.
Finally, the gravitational force on the object ball is represented by $\vec{F}_g$.
During the collision, the normal force on the object ball is given by
\be{F'N}
\vec{F}'_N  \equiv F'_N \hat z ; \qquad
F'_N=( M g - f \sin\theta) \Theta(\theta_c -\theta).
\ee
The object ball stands still before the collision and then starts moving. 
As discussed in the introduction, the object ball, being supported by the soft fibers of the table, experiences motion only influenced by these fibers during the short moment of collision. 
In this brief moment, the displacement of the object ball is extremely small, to the extent that it is difficult to consider it moving through other fiber materials. 
Therefore, during the collision, the frictional force between the object ball and the table is given by static friction.
The maximum static friction coefficient\footnote{To measure the static friction coefficients between the table and billiard balls, we perform an experiment. Three balls were bonded together using adhesive to prevent sliding. A weight was then used to gradually pull the set of three balls until slipping occurred. The force required to initiate sliding was measured, and from these experiments, the maximum static friction coefficient ($\mu_{\rm s}$) was determined. The obtained values help characterize the static frictional interaction between the table and billiard balls. } is approximately
$$
 0.2 < \mu_{\rm s}  < 0.4.
$$
This value, being relatively larger than the coefficients of kinetic friction or rolling friction, cannot be ignored, and its influence is notable in the context of the interaction with the table. 
The relative velocity, $\vec V_= = - \hat z \times \hat z \times (\vec V+ R \hat z \times \vec \Omega)$, of the contact point of the object ball with the table will be given by the sum of the translational velocity and the rotational velocity of the object ball.
The rotational motion of the object ball is developed by the friction between the two balls, which can be considered negligible compared to the translational motion.
In addition, consulting the result of the perfectly elastic collision, the motion of the object ball mainly directed to the $x$-direction.
The friction force acts in the opposite direction of relative motion. Therefore, for the friction force $\vec{f}'_g,$ it can be approximated as  
\bea
\vec{f}'_g  \approx - f'_g \hat V_=,
\eea
where $\hat V_= $ is defined as $\hat V_= \equiv \frac{\vec{V}_=}{|\vec{V}_=|} \approx \hat x.$
The magnitude of the friction force between the objective ball and the table is given by 
$$
f'_g = \mu_{\rm s} F'_N,
$$​​
where $F'_N$  is the normal force supporting the object ball. 
If we neglect gravity, this frictional effect is only significant when there is topspin of the cueball.
 
As discussed in the introduction, if the cue ball is moving rapidly $(U_i \gtrsim 1\,{\rm m/sec})$and collides with the objective ball at an angle $\theta \nsim 0$, the vertical component of the frictional force becomes significantly larger than the gravitational force. Therefore, in this case, it is permissible to approximate the normal force as
\be{FN2}
F_N \approx f \sin\theta \, \Theta(\theta), \qquad  
F'_N \approx - f \sin\theta \,\Theta(-\theta) .
\ee
The subsequent calculations will utilize this approximation.

\vspace{.2cm}
Following the discussion above, the equations of motion for the changes in the center velocities of the cue ball and the objective ball can be written as:
\bea
M\frac{d\vec{U}}{dt} &=&- \vec{F} +\vec{f}_t + \vec{F}_g + \vec{F}_N 
\approx F_\perp\left[- \hat x
- \mu \cos \theta \,\hat y - \mu \sin \theta \Theta(-\theta) \hat z \right]
, \nn \\
M\frac{d\vec{V}}{dt} &=& \vec{F} +\vec{f}'_t + \vec{F}_g + \vec{F}'_N 
\approx F_\perp \left[ \left(1+ \mu_{\rm s} \mu \sin\theta \, \Theta(-\theta) \right)\, \hat x + \mu \cos \theta \, \hat y +\mu \sin\theta \, \Theta(\theta) \hat z  \right]  ,
 \label{dU dt}
\eea
where $F_\perp$ is the perpendicular component of the collision force.
The rotational motion equations for the changes in the angular velocities $\vec{\omega}$ and $\vec{\Omega}$ of the cue ball and the objective ball are given by:  
\bea
I\frac{d\vec{\omega}}{dt} &=& \vec{\tau}_{-f} + \vec{\tau}_{f_g} 
	= R \hat{x} \times (-\vec{f}) - R \hat{z} \times \vec{f}_g
	\approx R \mu F_\perp
		 (\sin \theta \,\hat y- \cos\theta \,\hat z) , \nn \\
I\frac{d\vec{\Omega}}{dt} &=&  \vec{\tau}_{f} + \vec{\tau}_{f'_g} 
	= (-R \hat x) \times \vec{f}  - R\hat z \times \vec{f}'_g 
	\approx \mu RF_\perp \left\{ \sin \theta\left[1 + \mu_{\rm s} \Theta(-\theta)\right] \,\hat y-  \cos\theta \,\hat z  \right\}
	, 
\label{dom dt}
\eea
where,  $I= \nu MR^2$represents the rotational inertia of the ball with $\nu = 2/5$ for a uniformly dense sphere. 
In these equations, the first/second term represent the torques due to the collision between the balls and due to the friction between the balls and the table, respectively. 
As observed, neglecting the friction with the table ($\mu_{\rm s} \to 0$), would imply equal changes in the angular velocities of both balls. However, as discussed earlier, the value of the static friction coefficient cannot be easily dismissed.

\vspace{.2cm}
The frictional forces between the cue ball and the object ball act in the opposite direction of the relative velocity at the point of collision. To calculate the relative velocity at the point of collision for each ball, let's denote the displacement from the center of the cue ball to the collision point as $\vec{R}_c = R \hat x$. 
The velocity at the collision point of the cue ball is given by $\vec{U}+ \vec{\omega}\times \vec{R}_c$, and for the objective ball, it is $\vec{V}+ \vec{\Omega} \times \vec{R}_o$, where $\vec{R}_o = -  \vec{R}_c$.  
Therefore, the relative velocity of the objective ball with respect to the cue ball at the two contact points is given by
$$
\vec{v} = \vec{V}-\vec{U}-\vec{R}_o\times \vec{\Omega}  +\vec{R}_c\times \vec{\omega}.
$$                 
Conversely, the relative velocity of the cue ball with respect to the objective ball at the contact point is  $-\vec{v}$.
The initial values of the relative velocities are given by
\be{vi}
v_x(0) = v_\perp(0) = v_{\perp 0} = - U_i \cos \phi, \qquad
v_y (0) = -U_i \sin\phi - R\omega_z(0), \qquad
v_z(0) = R \omega_y(0) .
\ee
Here, the initial directional angle $\theta_0$ for the frictional force satisfies $\tan \theta_0 = v_{z0}/v_{y0}$, and
\be{v p 0}
v_\parallel(0) = v_{\parallel 0} 
= \sqrt{ [U_i \sin\phi + R\omega_z (0)]^2 +
	 R^2\omega_y^2(0)}    .
\ee
Because the friction acts along the opposite direction to the relative velocity, the angle satisfies
\bea
\cos\theta_0 = - \frac{U_i \sin\phi/R+ \omega_{z0}}{ \sqrt{(U_i \sin\phi/R+ \omega_{z0})^2+ \omega_{y0}^2 } },  \quad
\sin\theta_0 =  \frac{-\omega_{y0}}{ \sqrt{(U_i \sin\phi/R-\omega_{z0})^2+ \omega_{y0}^2 } }. 
\label{f direction}
\eea
The time-dependent change in the relative velocity is then given by
\bea
\frac{d\vec{v}}{dt} &=&\frac{F_\perp}{M} \left[ \Big( 2+ \mu_{\rm s} \mu \sin\theta \, \Theta(-\theta) \Big) \hat x  +\frac{2(1+\nu)}{\nu}  \mu \cos \theta\, \hat y
+\frac{2 +\nu+ \mu_{\rm s} \Theta(-\theta)}{\nu }  \mu \sin\theta \, \hat z\right] ,
\label{dv/dt}
\eea
where we use Eqs.~\eqref{dU dt} and \eqref{dom dt}.

Let's denote the normal (perpendicular to the collision plane) impulse exerted on the balls during the collision as
\be{p perp}
p\equiv p_\perp .
\ee
This value monotonically increases during the collision. 
We will use this impulse in place of time. 
Define the impulses in the directions perpendicular and parallel to the collision plane during the short time $dt$ as 
\be{dp}
dp \equiv dp_\perp = F_\perp dt = M dV_\perp, \quad
dp_\parallel = F_\parallel dt = f dt = \mubb F_\perp dt = M dV_\parallel = \mubb dp_\perp ,
\ee
respectively.
Now, the equations of motion \eqref{dU dt} can be expressed by multiplying both sides by
$dt$ and using the defined expressions as follows:
\bea
Md\vec{U} 
&=& dp\left[- \hat x
- \mu \cos \theta \,\hat y - \mu \sin \theta \Theta(-\theta) \hat z \right], \nn \\
Md\vec{V} &=&   dp \left[ \left(1+ \mu_{\rm s} \mu \sin\theta \, \Theta(-\theta) \right)\, \hat x + \mu \cos \theta \, \hat y +\mu \sin\theta \, \Theta(\theta) \hat z  \right] 
\label{dU dV}
\eea  
Regarding the change in angular velocity in \eqref{dom dt}, when expressed in terms of impulse, it becomes:
\be{dom}
Id\omega_y \approx \mu R\sin\theta \, dp  
\approx \frac{I d\Omega_y}{1+ \mu_{\rm s} \Theta(-\theta)},
\qquad
Id\omega_z \approx -\mu R dp_\perp \cos\theta \approx Id\Omega_z  .
\ee
Here, $\approx$ indicates an approximation neglecting gravity and friction between the ball and the table.

The change in relative velocity with respective to the impulse can be obtained as follows:
\bea \label{dv dp}
d\vec{v} =\frac{dp_\perp}{M} \left[ \Big( 2+ \mu_{\rm s} \mu \sin\theta \, \Theta(-\theta) \Big) \hat x  +\frac{2(1+\nu)}{\nu}  \mu \cos \theta\, \hat y
+\frac{2 +\nu+ \mu_{\rm s} \Theta(-\theta)}{\nu }  \mu \sin\theta \, \hat z\right] 
\eea
Here, recalling that
\be{v rel}
v_y \equiv - v_\parallel \cos \theta, \qquad 
v_z = - v_\parallel \sin\theta
\ee
and utilizing  $v_\parallel d\theta = \sin \theta \, dv_y - \cos \theta \, dv_z$, we find
\be{d theta}
v_\parallel d\theta = \frac{\mu \sin 2\theta \, dp_\perp}{2M}\frac{\nu -\mu_{\rm s} \Theta(-\theta)}{\nu} .
\ee
As observed from this equation, the angle $\theta$, which indicates the direction in which frictional force is acting, changes over time. 
Because $\nu (=2/5) > \mu_{\rm s}$, $d\theta$ becomes negative when $\theta < 0$, causing the angle $\theta$ to decrease.
The change in the angle becomes zero at $\theta =0, \pm \pi/2$ and reaches its maximum at $\theta = \pm \pi/4$, happening in the direction of increasing $|\theta|$.

In both cases, there is no event during the collision that reverses the sign of the angle $\theta$.
Moreover, as evident from the equations of motion, the friction coefficient $\mu_{\rm s}$always comes with a step function $\Theta(-\theta)$, and the sign of $\theta$ does not change during the motion. 
Therefore, for simplicity, we will denote $\mu_{\rm s} \Theta(-\theta)$ as $\mu_{\rm s}$ later in this work.

\section{Collision process }

In this section, we aim to provide a detailed chronological description of the process that occurs at the moment when billiard balls collide. 
To illustrate the sequence of physical phenomena during the collision process, we focus on the impulse $p\equiv p_\perp$ acting in the perpendicular direction to the collision plane. 
If there is no adhesive force between the balls on the collision plane, the impulse $p$ monotonically increases over time during the collision process. 

\begin{enumerate}
\item {\bf Sliding state:}
The moment the cue ball reaches the object ball, both balls enter a state of sliding against each other. 
At the moment of collision, the cue ball simultaneously pushes the object ball vertically with a speed of $U_i\cos\phi$ while sliding on the surface of the ball with a speed of $U_i\sin\phi$. 
This sliding is gradually slowed down by friction, and if the friction coefficient is denoted as $\mubb$, the impulse due to frictional force is given by
 \be{f parallel}
 dp_\parallel = \mubb dp
 \ee
Here, we assume that the friction coefficient is independent of the sliding speed or vertical pressure, which is valid as long as the sliding speed on most surfaces is not too fast. 
In collisions between balls, the friction coefficient $\mubb$ is not large, making it difficult to enter a stick state where sliding stops.
During sliding, changes in the horizontal and vertical components of relative velocity are given by
\be{dv/dp}
\frac{dv_y}{dp} 
= -\frac{2\mubb }{M} \frac{\nu+1}{\nu} \frac{v_y}{v_\parallel} , \qquad
\frac{dv_z}{dp} = -\frac{\mubb }{M} \frac{\nu+2+\mu_{\rm s} }{\nu} \frac{v_z}{v_\parallel}  , \qquad 
\frac{dv_\perp}{dp} = \frac{2+\mu\mu_{\rm s} \sin\theta }M
\ee
As mentioned earlier, $\mu_{\rm s}$ represents $\mu_{\rm s} \Theta(-\theta)$. The last equation indicates that for $\theta> 0$, $v_\perp$ linearly varies with the vertical impulse. On the other hand, for $\theta < 0$, $v_\perp$ no longer linearly varies with $p$.

Due to the step function, the equation \eqref{dv/dp} needs to be solved separately for each case where $\theta$ is negative or positive. Fortunately, the results for the case of $\theta> 0$ can be obtained by solving the case of $\theta< 0$ first and then taking the limit, $\mu_{\rm s} \to 0$. Therefore, it is sufficient to solve the case of $\theta< 0$. In this case, $v_y < 0$, $v_z > 0$. Integrating the first two equations yields a conserved quantity during the collision process:
\be{bar v}
\bar v\equiv v_z^{-\frac{2(1+\nu)}{\mu_{\rm s}-\nu}} |v_y|^{ \frac{2+\nu+\mu_{\rm s}}{\mu_{\rm s}-\nu}} 
	= |v_y| \left| \frac{v_y}{v_z}\right|^{\frac{2(1+\nu)}{\mu_{\rm s}-\nu}}
	=v_z \left| \frac{v_y}{v_z}\right|^{\frac{2+\nu +\mu_{\rm s}}{\mu_{\rm s}-\nu}} = \mbox{constant} .
\ee
Thus, the magnitude of the component parallel to the collision plane of the relative velocity is given by
$
 v_\parallel = |v_y| \sqrt{1+ x_y^{1/a_y } } 
 = |v_z| \sqrt{1+ x_z^{1/a_z}} .
$
Here,
\be{ay az}
a_y \equiv -\frac{1+\nu}{\nu-\mu_{\rm s}}, \quad 
a_z \equiv \frac{2+\nu +\mu_{\rm s}}{2(\nu-\mu_{\rm s})} ,
\ee
and
\be{x yz}
x_y \equiv \left|\frac{v_y}{\bar v}\right| = \big(\tan^2\theta\big)^{a_y},
\qquad
x_z \equiv  \left|\frac{v_z}{\bar v}\right| = (\tan^2\theta)^{-a_z} .
\ee

From the initial values~\eqref{vi}, it follows that
$$
\bar v = |R \omega_y(0)| \left| \frac{R\omega_y(0)}{U_i \sin\phi + R\omega_z(0)} \right|^ {2 a_z}
$$
Using this, the first two equations in Eq.~\eqref{dv/dp} become
\be{dv/dp 2}
\sqrt{1+ x_k^{1/a_k } } dx_k
	= -\frac{\mubb \beta_k }{m \bar v}  dp , \qquad k= y, z ,
\ee
where
\be{beta yz}
\beta_y = \frac{\nu+1}{\nu}, \qquad
\beta_z = \frac{\nu+2+\mu_{\rm s}}{2\nu}
\ee
The differential equations \eqref{dv/dp 2} can be integrated using hypergeometric functions:
\be{vp2}
\mathfrak{V}_k = \mathfrak{V}_{k0} 
	- \frac{\mubb \beta_k p}{m \bar v} .
\ee
Here,
\be{VV2}
\mathfrak{V}_k \equiv  F_{a_k}( x_k)
\equiv \int^{x_k} \sqrt{1+ x^{1/a_k}} dx 
	= x_k\, {}_2F_1(-1/2,a_k, 1+a_k, -x_k^{1/a_k}), 
\ee
and $\mathfrak{V}_{k0} $ ($k= y, z$) is determined by the relative velocity components $v_y(0)$, $v_z(0)$ at $p=0$. This function monotonically increases in the region, $x>0$. Finally, considering the motion perpendicular to the collision plane from the last equation of \eqref{dv/dp}, we get
\be{v perp}
v_\perp -v_\perp(0)= \frac{p}{m} + \frac{\mu \mu_{\rm s}}{M} \int^p \sin\theta dp
= \frac{p}{m} + \frac{ \mu_{\rm s} [v_z-v_z(0)]}{2\beta_z} .
\ee
Here, $m \equiv M/2$ and Eq.~\eqref{int cos} in appendix is used.

The results for positive $\theta$ can be obtained by taking the limit as $\mu_{\rm s} \to 0$. For example, in the case of relative velocity perpendicular to the collision plane, equation \eqref{v perp} simplifies to $v_\perp = p/m$. If $v_\parallel \approx 0$, even with a small friction coefficient, a stick state can be entered in a short time, but this paper does not consider such possibility.

\item {\bf Maximally compressed state:}
As the relative vertical velocity decreases, the two balls are compressed, and energy is accumulated. Let's denote the vertical impulse at the moment when this vertical compression stops, i.e., when $v_\perp = 0$, as $p_\perp = p_c$. At that moment, the relative vertical velocity is $0$, and the following equations hold:
\be{v perp c}
v_\perp(p_c) = v_{\perp 0} + \frac{p_c}{m} 
	+\frac{ \mu_{\rm s}[v_z(p_c)-v_z(0)]}
			{2\beta_z}  =0, \quad
\mathfrak{V}_k(p_c) =\mathfrak{V}_k(0) - \frac{\mubb \beta_k p_c}{m \bar v} .
\ee
By solving the first equation and the second equation for $k=z$, we can determine the value of $p_c$. In the case of $\theta > 0$, taking the  $\mu_{\rm s} \to 0$ limit yields the result that the vertical impulse up to the moment of maximum compression is $p_c = -mv_{\perp 0} = m U_i \cos \phi$.

The total energy absorbed during the compression process can be obtained by integrating the force acting in the vertical direction over the displacement. 
If the vertical component of the relative velocity varies linearly with the impulse during compression, the magnitude of the work done as the vertical impulse increases from 0 to $p_c$ can be expressed using equations \eqref{dv/dp}, \eqref{v perp}, and \eqref{VV2}: 
\bea \label{W perp c}
W_{\perp}(p_c) &=& \int_0^{p_c} v_\perp dp_\perp
= v_{\perp0} p_c+ \frac{p_c^2}{2m} +
 	\frac{\mu_{\rm s}}{2\beta_z}
	\left[- v_{z0}p_c + \int_0^{p_c} dp v_z \right] \nn \\
&=& v_{\perp0} p_c+ \frac{p_c^2}{2m} -
 	\frac{\mu_{\rm s} }{2\beta_z}
	\left[v_{z0}p_c 
	+ \frac{M \bar v^2}{4\mu}
 		\frac{\left[ F_{ 2a_z } (x_z^2)
			-F_{ 2a_z } (x_z^2(0)) \right] }{\beta_z} 
  \right] 	.
\eea
In the case of $\theta > 0$, taking the limit as $\mu_{\rm s} \to 0$ simplifies the result to the first two terms of equation \eqref{W perp c}. 

\item {\bf Restoration process:}
At the moment when the vertical velocities of the cue ball and the object ball become equal, both balls stop compressing and enter the restoration process. 
During this period, as the shapes of the two balls return to their original state, the elastic energy accumulated during the compression process is utilized to push each other away. 
Let's denote the final value of the vertical impulse obtained during the restoration process as $p_f$, and explore the kinetic energy restored during this process.
Since both balls are in a sliding state during the restoration process, the vertical and horizontal impulses satisfy a similar form of equation as in \eqref{v perp c}. Integrating this equation up to $p_c \sim p_f$ determines the final state.

The final horizontal relative velocity is given by \eqref{vp2}:
\be{vf paral}
\mathfrak{V}_k(p_f) = \mathfrak{V}_k (p_c) - \frac{\mu \beta_k}{m \bar v}(p_f-p_c) = \mathfrak{V}_{k 0} - \frac{\mu \beta_k}{m \bar v} p_f .
\ee
Therefore, the current assumption that the friction is small enough not to reach a stick state still holds if $\frac{\mu \beta_k}{m} p_f < \mathfrak{V}_{k 0}$ for $k=y,z$. As it is still in a sliding state, the vertical relative velocity also satisfies equation \eqref{v perp}. Thus,
\be{vf perp}
v_{\perp f} =v_\perp(0)
+\frac{p_f}{m} + \frac{ \mu_{\rm s}}{2\beta_z} [v_{zf}-v_z(0)]
=\frac{p_f-p_c}{m} + \frac{ \mu_{\rm s}}{2\beta_z} [v_{zf}-v_z(p_c)]
,
\ee
where, in the second equality, the first equation of \eqref{v perp c} is used. If $\theta$ is positive, $v_\perp(p_f) = \frac{p_f-p_c}{m}$ can be expressed more succinctly.

The elastic energy restored during the restoration process can be calculated using the above equation:
\bea \label{W perp f}
W_{\perp} (p_f) - W_{\perp} (p_c) &\equiv & \int_0^{p_c} v_\perp dp_\perp= 
\int_{p_c}^{p_f} \left[\frac{p-p_c}{m} 
	+ \frac{ \mu_{\rm s}}{2\beta_z} [v_{z}-v_z(p_c)]
\right] dp\nn \\
&=&  \frac{(p_f - p_c)^2}{2m }
-\frac{ \mu_{\rm s}}{2\beta_z}
\left[ v_z(p_c)( p_f-p_c)   + \frac{M \bar v^2}{4\mu \beta_z}
	\Big(F_{2a_z} (x_{zf}^2) - F_{2a_a} (x_{zc}^2) \Big) 
\right].
\eea
\end{enumerate}

From the above results, the ratio of absorbed elastic energy during the compression process to the restored energy can be related to the square of the (energy) coefficient of restitution, given by:
\be{e*2}
e_*^2 = - \frac{W_\perp(p_f) - W_\perp(p_c)}{W_\perp(p_c)} .
\ee
Here, $e_*$ represents the energetic coefficient of restitution between the cue ball and the object ball~\cite{Stronge2018}. 
From this equation, one can write the value of $p_f$ in terms of the initial value. 
Generally, although this equation presents a complex relationship, in the case of $\theta > 0$, the right side of this equation is simplified, and $e_*^2 = \left(\frac{p_f}{p_c}-1\right)^2$ is given. Therefore, in this case,
$p_f = (1+ e_*) m U_i \cos \phi .$                     
In general, $p_f$ has a correction term proportional to $\mu_{\rm s}$ in addition to the above equation:
\be{pf}
p_f = (1+ e_*) m U_i \cos \phi + \mu_{\rm s} \delta p_f .
\ee

Using the above results to find the post-collision velocity of the cue ball, with the help of \eqref{dU dV} and \eqref{int cos}, we have: 
\bea \label{Uf}
U_{\perp f} &=& U_{\perp i} -\frac1M \int_0^{p_f} dp_\perp 
= \frac{1-e_*}{2}U_i \cos \phi - \frac{\mu_{\rm s} \delta p_f}{M}, \nn \\
U_{yf}&=& U_{yi} - \frac{\mu}M\int_0^{p_f} \cos \theta \, dp
 = U_i \sin\phi -  \frac{1}{2\beta_y} [v_y(p_f) - v_y(0)],  \nn \\
 U_{zf} 
& \approx &U_{zi} -  \frac{\mu\Theta(-\theta)}{M}\int_0^{p_f} \sin\theta \, dp
= - \frac{1}{2\beta_z}[ v_z(p_f) - v_{z}(0)]\Theta(-\theta),
\eea
where $v_k(p_f)$ is obtained from equation \eqref{vf paral}, and $v_k(0)$ is obtained from equation \eqref{vi}. Also, $U_{zi} = 0$. The post-collision velocity of the object ball is given by:
\bea \label{Vf}
V_{\perp f} &=& \frac1M\int (1+\mu\mu_{\rm s} \sin\theta)dp_\perp = \frac{1+e_*}{2} U_i\cos \phi 
	+ \frac{\mu_{\rm s} \delta p_f}{M}
	+ \frac{ \mu_{\rm s} } {2\beta_z} \left[ v_z(p)-v_z(0) \right] , 
\nn \\
V_{yf} &=& \frac{\mu}M\int \cos\theta \, dp
	=  \frac{1}{2\beta_y} [v_y(p_f) - v_y(0)]
, \nn \\
V_{zf} &=&  \frac{\mu \Theta(\theta)}M \int  \sin\theta \, dp
= \frac{1}{2\beta_z}[ v_z(p_f) - v_{z}(0)] \, \Theta(\theta).
\eea
In the $y$-direction, the law of conservation of momentum holds, but it does not hold in the $x$ and $z$-directions. 
This phenomenon occurs due to the counteracting effect of vertical forces, neutralizing the impact of frictional forces.

The angular velocities of the two balls after the collision become
\bea \label{dom2}
\omega_{\perp f}- \omega_{\perp i} &=&0 =\Omega_{\perp f}  , \nn \\
\omega_{yf}- \omega_{yi} 
&=& \frac{\mu R}{I} \int dp \sin\theta 
=  \frac{M R}{I} \frac{ v_{zf}- v_{zi}  }{2\beta_z}
 = \frac{\Omega_{yf}}{1+ \mu_{\rm s} \Theta(-\theta)} , \nn \\
\omega_{zf}- \omega_{zi} 
&=& -\frac{\mu R}{I}\int dp \cos\theta  
= -\frac{MR}{I} \frac{ v_{yf}- v_{yi}}{2\beta_y } = \Omega_{zf} ,
\eea
as derived from \eqref{dom}.
Utilizing the above results to compare the energy before and after the collision, one may notice that the conservation of kinetic energy does not hold when there is friction and the coefficient of restitution is not 1.

\section{Linear approximation for the motion of the cue ball and object ball after the collision}

In the previous section, we discussed the collision between two balls in a general form. 
However, in the case of a collision between billiard balls, the magnitude of the coefficient of friction $\mubb$ between the balls is not large, resulting in a small change in the relative velocity in the direction parallel to the collision plane. 
In such cases, it is sufficient to use linear approximation by expanding the function $\mathfrak{V}_k$ with $k=y,z$ around the initial values.

Firstly, from the definition of the function $F_a$ in Eq. \eqref{VV2}, we obtain the following expansion:
\be{F series}
F_a(x) - F_a(x_0) \approx \sqrt{1+ x_0^{1/a}} (x-x_0)\left[1
	+ \frac1{2a} \frac{x_0^{1/a-1} }{1+ x_0^{1/a} } (x-x_0)+\cdots \right].
\ee
We have described up to quadratic approximation, as it is necessary for expanding the coefficient of restitution formula \eqref{e*2} to the linear approximation in terms of the friction coefficient $\mu$. Conversely, we can also use
\be{x series}
x- x_0 \approx \frac{F_a(x) - F_a(x_0) }{\sqrt{1+ x_0^{1/a}} }
	\left[1- \frac1{2a} \frac{x_0^{1/a-1} }{\left(1+ x_0^{1/a} \right)^{3/2} } \big(F_a(x) - F_a(x_0) \big)+\cdots \right] .
\ee

First, we calculate $v_{zf} - v_{zi}$, which is necessary to obtain the value needed for expanding the coefficient of restitution formula, up to quadratic approximation in terms of $\mu$.
 Using Eqs.~\eqref{vp2} and \eqref{x series}, we get
$$
|v_z|- |v_{z0}| \approx - \beta_z \frac{\mu p |\sin\theta_0|}{ m} 
 \left[1-\frac{\mu_{\rm s}-\nu}{2\nu}  \cos^2\theta_0
		\frac{\mu p}{m v_{\parallel 0}} 
	 	 +\cdots \right] .
$$
The calculation process uses Eq. \eqref{v rel} and Eq. \eqref{x yz} also. 
If $\theta < 0$, $v_z>0$, so removing the absolute value on the left-hand side is not a problem. On the right-hand side, the $-$ sign can be absorbed into $\sin \theta_0$. If $\theta> 0$, the entire left-hand side gets a $-$ sign, allowing us to cancel the absolute value and the $-$ sign on the right. Therefore, the following generally holds:
\be{delta vz}
 v_z- v_{z0} 
 = \beta_z \frac{\mu p \sin\theta_0}{ m} 
 \left[1+\frac{\nu-\mu_{\rm s}}{2\nu}  \cos^2\theta_0
		\frac{\mu p}{m v_{\parallel 0}} 
	 	 +\cdots \right].
\ee
In future calculations, $v_{yf}- v_{yi}$ only needs to be approximated up to the first order in $\mu$. Using Eq. \eqref{vp2} and $v_y< 0$, we get:
\be{delta vy}
v_y-v_{y0} =- (|v_y|- |v_{y0}|) \approx -\bar v\frac{\mathfrak{V}_y - \mathfrak{V}_{y0} }{\sqrt{1+ x_{y0}^{1/a_y}} }+ \cdots
 =  \beta_y \frac{\mu p \cos\theta_0}{ m} + \cdots
\ee

Then, let's solve for $p_c$ using Eq. \eqref{v perp c} by determining it around the initial value. From the first equation, we get:
$$
v_\perp(p_c) = v_{\perp 0} + \frac{p_c}{m} 
	+ \frac{\mu \mu_{\rm s} p \sin\theta_0}{2 m} 
 \left[1-\frac{\mu_{\rm s}-\nu}{2\nu}  \cos^2\theta_0
		\frac{\mu p}{m v_{\parallel 0}} 
	 	 +\cdots \right]=0
$$
Therefore, the following holds:
\be{pc}
v_{\perp 0} = - \frac{p_c}{m} \left[1
+ \frac{\mu \mu_{\rm s} \sin\theta_0}{ 2}\right] +\cdots 
\ee

\vspace{.3cm}
Now, using the above results and the coefficient of restitution formula \eqref{e*2}, let's determine $p_f$. First, during the contraction, the elastic energy accumulated in the ball can be obtained from \eqref{W perp c}. Using \eqref{delta vz} and \eqref{pc}, we get:
\bea
W_\perp(p_c) &=&  v_{\perp0} p_c+ \frac{p_c^2}{2m} 
	+\frac{\mu_{\rm s} \Theta(-\theta)}{2\beta_z}
	\left[- v_{z0}p_c - \frac{M \bar v^2}{4\mu \beta_z} 
		\left[ F_{ 2a_z } (x_z^2)
	-F_{ 2a_z } (x_z^2(0)) \right]
  \right] 	\nn \\
&\approx& -\left[1
	+ \frac{\mu \mu_{\rm s}  \sin\theta_0 }{2} 
	\right]\frac{p_c^2}{2m} .
\label{Wc}
\eea
Here, $v_{z0}= - v_{\parallel 0} \sin\theta_0$ and if $\mu_{\rm s}\neq 0$, $\theta< 0$ is used.
The elastic energy recovered in the restoration process can be obtained from \eqref{W perp f}. Using a similar process as above, we get:
\bea
W_{\perp} (p_f) - W_{\perp} (p_c) 
&=&  \frac{(p_f - p_c)^2}{2m }-\frac{ \mu_{\rm s}}{2\beta_z}
\left[ v_z(p_c)( p_f-p_c)   + \frac{M \bar v^2}{4\mu \beta_z}
	\Big(F_{2a_z} (x_{zf}^2) - F_{2a_a} (x_{zc}^2) \Big) 
\right] \nn \\
&\approx&  \frac{(p_f - p_c)^2}{2m }\left[1 +\frac{\mu \mu_{\rm s}}{2} \sin\theta_c \right] .
\eea
Now, using \eqref{e*2}, we have:
$$
e_*^2 = \left(\frac{p_f}{p_c} -1\right)^2 
	\frac{2+ \mu \mu_{\rm s} \sin\theta_c}
		{2+ \mu \mu_{\rm s} \sin\theta_0}+ O(\mu^2) .
$$ 
Here, $\theta_c-\theta_0 \sim O(\mu)$, so we obtain:
\be{pf pc}
p_f \approx \left[1 +e_* + O(\mu^2)\right] p_c
\ee
In other words, up to the first-order approximation in $\mu$, the vertical impulse is determined by the coefficient of restitution. Therefore, using \eqref{pc} and \eqref{vi}, the final vertical impulse is given by:
\be{pf2}
p_f 
= (1+e_*) m U_i \cos\phi \left[1- \frac{\mu \mu_{\rm s} \sin\theta_0}{2} +\cdots \right].
\ee
Comparing this equation with \eqref{pf}, the impulse correction is:
\be{delta pf}
\delta p_f = -(1+e_*)m U_i \cos \phi \frac{\mu \sin \theta_0}2
\ee

Now, we write the velocities of the cue ball and the object ball after the collision given by Eqs. \eqref{Uf} and \eqref{Vf}:
\bea \label{Uf2}
\vec{U}_{ f} &\approx&
	U_i \cos \phi \left[\left(\frac{1-e_*}{2}+ \frac{\mu \mu_{\rm s}(1+e_*)\sin \theta_0}{4}\right)\hat x  \right. \nn \\
&& \left.+  \left(\tan\phi - \frac{\mu (1+e_*) \cos\theta_0}{ 2}\right) \hat y
	-\frac{\mu (1+e_*) }{ 2} \sin\theta_0 \Theta(-\theta) \hat z
	\right],  \\
\vec{V}_{f} &\approx& 
\frac{1+e_*}{2} U_i\cos \phi \left[\left(1
	+ \frac{\mu \mu_{\rm s} \sin \theta_0}{2}\right) \hat x 
	+ \mu \cos \theta_0 \, \hat y + \mu \sin \theta_0 \, \Theta(\theta_0) \, \hat z\right] 
\label{Vf2}
\eea
In the case where there is a draw, i.e., $\theta_0 > 0$, one can take the limit $\mu_{\rm s} \to 0$ in the above results. 
As seen from these equations, for $\theta_0 < 0$, the friction between the object ball and the ground reduces the speed in the direction perpendicular to the collision plane by the same amount for both the cue ball and the object ball.

Now, let's examine the changes in angular velocity for the cue ball and the object ball.
Firstly, there is no change in angular velocity component in the direction perpendicular to the collision plane. 
On the other hand, the angular velocity components along $y$ and $z$-directions change according to Eq. \eqref{dom2}:
\bea \label{omega}
\vec\omega_{f}- \vec\omega_{i} &=&
 \frac{\mu (1+e_*)  }{2\nu }\frac{U_i \cos\phi }{R} \left[ \sin \theta_0 \, \hat y - \cos \theta_0 \, \hat z\right], \nn \\
\vec\Omega_{f}  &=&  \frac{\mu (1+e_*)  }{2\nu }\frac{U_i \cos\phi }{R} 
	\left[(1+ \mu_{\rm s}) \sin \theta_0 \, \hat y - \cos \theta_0 \, \hat z\right]
\eea
This means that the object ball obtains a larger angular velocity with a higher initial horizontal speed and a larger friction coefficient. Additionally, we observe that the friction between the object ball and the ground only contributes to the change in the angular velocity of the object ball when there is a draw.

\section{Analysis of the results and discussions}
\begin{figure}[tbh]
\begin{center}
\begin{tabular}{c}
\includegraphics[width=.5\linewidth]{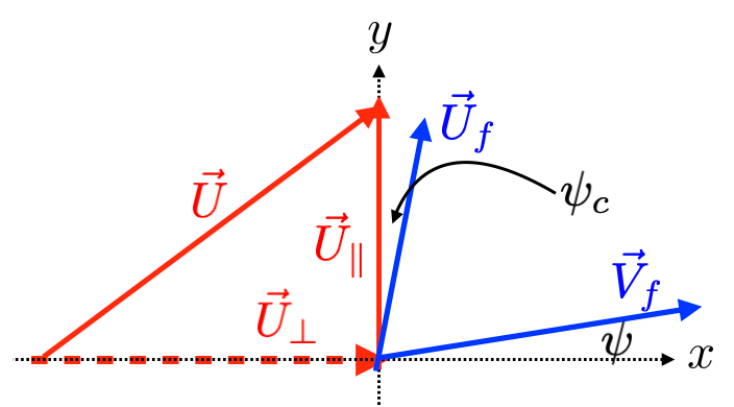} 
\end{tabular}
\end{center}
\caption{\footnotesize The velocity changes of the cue ball and the object ball before and after the collision (ignoring the component of velocity in the direction perpendicular to the table)}
\label{fig:vt3}
\end{figure}

In this work, we discusse the phenomenon that occurs when a cue ball, rotating on a plane, approaches and collides with a static object ball of equal mass. 
In the case of billiard balls, the coefficient of restitution is close to $1$, and the friction coefficient takes a value close to $0$, demonstrating a collision phenomenon close to a perfectly elastic collision. 
In this scenario, immediately after the collision, the trajectories of the cue ball and the object ball form a right angle with each other. The cue ball moves in the direction parallel to the collision plane, while the object ball moves in the direction perpendicular to the collision plane. 
This situation is illustrated in Figure~\ref{fig:vt3} using red arrows, where $\vec{U}_i$ represents the velocity of the cue ball before the collision, and $\vec{U}_\perp$ and $\vec{U}_\parallel$ represent the velocities of the object ball and the cue ball, respectively, after a perfectly elastic collision on the table.

In this paper, three effects which develop  deviations from the results of a perfectly elastic collision in the collision of balls are discussed. 
These three effects are: 1) the collision between balls is not perfectly elastic, 2) the effect of friction between balls during collision, and 3) the effect of friction between the ball and the table at the moment of collision. 

Before discussing the detailed results of all these effects, let's first consider the scenario where a cue ball with topspin collides head-on with an object ball, taking into account the presence of frictions, the topic we mentioned at the beginning of the introduction.
Firstly, assume that the cue ball has only rolling motion before the collision. 
Therefore, $\theta_0 = -\pi/2$. 
In a head-on collision, $\phi = 0$. 
To simplify the discussion, we assume a perfectly elastic collision, implying $e_* = 1$.
As discussed earlier, the effects of friction between the object ball and the table are valid only when there is topspin in the cue ball. Therefore, we interpret the friction coefficient $\mu_{\rm s}$ in Eqs.~\eqref{Uf2} and \eqref{Vf2} as $\mu_{\rm s} \Theta(-\theta_0)$, where $\Theta$ is the step function. Then, the velocities of the cue ball and the object ball after the collision are: 
$\vec{U}_{ f} \approx
	\mu U_i \left(-\frac{\mu_{\rm s}}{2}\hat x 
	+ \hat z
	\right), ~~
\vec{V}_{f} \approx 
 U_i\left(1- \frac{\mu \mu_{\rm s}}{2}\right) \hat x .$
If there is no friction between the balls ($\mu=0$), as expected, the cue ball comes to a stop, and the object ball moves with the same speed as the cue ball. 
However, when there is friction between the balls and between the object ball and the table, after the collision, the cue ball moves backward at a rate slower by the factor $\mu_{\rm s}/2$ compared to the speed at which it would jump into the air. 
Of course, this backward motion will soon turn into forward motion due to the friction when the cue ball collides with the table again under the influence of gravity. 
Similarly, the object ball also moves slower by the same amount compared to its speed without friction with the table. 
This result clearly demonstrates that the friction between the object ball and the table cannot be ignored in the collision of two balls.

\vspace{.2cm} 
Now, let's describe the detailed results of the collision. 
Firstly, the phenomena arising from the difference with a perfectly elastic collision, due to the absence of friction, are proportional to $(1-e_*)$. 
This manifests as the appearance of the vertical velocity $U_{\perp f}$ of the cue ball and the decrease of $V_{\perp f}$ of the object ball. 
The phenomena solely caused by the friction between the cue ball and the object ball are proportional to $\mu$. 
Both balls exhibit changes in the velocity components parallel to the collision plane.
The changes in angular velocities, as seen in Eq.~\eqref{omega}, confirm that the presence of friction between the balls is necessary for such changes to occur. In Figure \ref{fig:vt3}, the blue arrows illustrate the trajectories of the cue ball and the object ball immediately after the collision.

Ignoring the motion perpendicular to the billiard plane and comparing the results with a perfectly elastic collision, the differences in the speed and the direction of the object ball are proportional to $(1-e_*)$ and $\mu$, respectively.
Let's examine the post-collision motion of the object ball \eqref{Vf2} using linear approximation. The object ball's speed is given by:
 $$
 V_f \approx \frac{1+e_*}2
 	 U_i \cos \phi \left(1+\frac{\mu\mu_{\rm s} \sin \theta_0}2\right)
 $$   
If we interpret $(1+e_*)/2$ as $1-(1-e_*)/2$, we can understand how it differs from a perfectly elastic collision. 
The velocity correction due to the friction coefficient is proportional to $\mu\mu_{\rm s}$ and exists only when $\theta$ is negative (in the presence of the cue ball's topspin).

As seen in Eq.~\eqref{Uf}, without friction, the object ball moves along the $x$-direction, perpendicular to the collision plane.
 If there is friction, the object ball's velocity has components in both the $y$-direction and the vertical upward $z$-direction parallel to the collision plane. 
The magnitude of the upward velocity component is given by:
$$
V_{f z} = \frac{\mu(1+e_*)}{2} U_i \cos \phi \, \sin \theta_0
	\Theta(\theta_0)
$$
This vertical motion is linearly proportional to the friction coefficient and occurs only when the cue ball has an angular velocity about the $y$-axis, i.e., there is topspin. 
The directional difference is denoted by $\psi$ in Figure \ref{fig:vt3} and is approximately given by:
\be{psi}
\psi \approx \tan ^{-1} \left(\mu \cos \theta_0\right)
\ee
This angle is maximized when there is no initial rotational angular velocity about the $y$-axis, and for a given friction coefficient according to Eq.~\eqref{mu1}, it varies by approximately $1.7^\circ$ to $4.8^\circ$. 
As seen, the primary cause for the object ball deviating from the direction perpendicular to the collision plane is the frictional force between the two balls during the collision.

When two balls of equal mass undergo a frictionless perfectly elastic collision, the cue ball moves in the $y$-direction after the collision. 
Neglecting the vertical upward motion, the motion of the cue ball, as seen from the first equation in \eqref{Uf2}, shows two changes. 
First, due to the effect of friction, the speed in the $y$-direction parallel to the collision plane decreases. 
Second, as the collision is not perfectly elastic, the cue ball gains a velocity component in the $x$-direction perpendicular to the collision plane. 
As a result, the cue ball deviates from the $y$-direction by an angle $\psi_c$ as depicted in Figure \ref{fig:vt3}, given by:
\be{psi w}
\psi_c \approx \tan^{-1} \left[\cot\phi\left( \frac{1-e_*}2+\frac{\mu \mu_{\rm s}(1+e_*)\sin \theta_0}4\right)\right] .
\ee
If the cue ball engages in a perfectly elastic collision ($e_* = 1$) with the object ball and has topspin (i.e., $\theta < 0$), the cue ball's direction deviates from the perpendicular direction to the collision plane because of the frictions. 
On the contrary, if there is no topspin, the direction of the cue ball is towards the $y$-axis. 
If a perfectly elastic collision does not occur, the cue ball's direction is determined by the combined effects of $(1-e_*)$ and the topspin effect through the friction between the object ball and the table.
Calculating the angular deviation using the friction coefficient of $0$ and the range of restitution coefficient given by Eq.~\eqref{e*}, at $\phi = 45^\circ$, for $e_* = 0.92$, there is an angular deviation of approximately $2.3^\circ$, and for $e_* = 0.98$, an angular deviation of about $0.57^\circ$. At $\phi = 30^\circ$, the deviations are approximately $4.0^\circ$ and $1.0^\circ$, respectively. For $\phi = 10^\circ$, an angular deviation of around $13^\circ$ is observed for $e_* = 0.92$. These deviations are substantial and cannot be ignored.
As the angle φ approaches 0 degrees (meaning closer to a head-on collision), the speed of the cue ball significantly decreases, and its direction is strongly influenced by the effect of friction.

As time progresses after the collision, the angle formed by the centers' velocities of the two balls changes due to the rotational motion. 
The post-collision trajectory of the balls is determined by the velocities/angular velocities resulting from the collision and the friction with the table.
 If there is angular velocity around the axis of the center's velocity of the ball, the ball follows a curved path. 
In the case of the object ball, the rotation induced by friction is not significant, and bending is not visibly apparent. 
However, the cue ball, depending on the situation, can exhibit significant curvilinear motion.
Further research is needed to investigate these trajectories in detail.

The results of this study contribute to the understanding of ball movements and collisions in billiards, providing insights into strategies and gameplay. Additionally, this research can offer valuable information in the field of physics education and experiments.

%
\section*{Acknowledgment}
This work was supported by the National Research Foundation of Korea grants funded by the Korea government RS-2023-00208047 and Korea National University of Transportation Industry-Academy Cooperation Foundation in 2023.

\appendix
\section{Integration of the trigonometry functions}
We attach an integration formula which is used frequently in the text:
\bea \label{int cos}
\int_0^{p} dp \cos\theta &=& \int \frac{-v_y}{v_\parallel} \frac{dp}{dv_y} dv_y 
= \frac{M}{2\mu \beta_y} \left[ v_y(p)-v_y(0) \right]
 \\
\int_0^{p} dp \sin\theta &=& \int \frac{-v_z}{v_\parallel} \frac{dp}{dv_z} dv_z 
= \frac{M}{2\mu \beta_z} \left[ v_z(p)-v_z(0) \right] .
\nn
\eea
In deriving this formula, we us Eq.~\eqref{dv/dp}.


\begin{thebibliography}{99}

\bibitem{Dave}
``Amazing BILLIARDS PHYSICS in Super Slow Motion", 
https://www.youtube.com/watch?v=NWkX9JCWCK0  (2021).

\bibitem{Coriolis}
Coriolis, ``Théorie mathématique des effets du jeu de billard," (1835);
``Mathematical Theory of Spin, Friction, and Collision in the Game of Billiards'', English translation, by David Nadler, ISBN: 0-9771671=0-0 (2005). 

\bibitem{Moore}
A. D. Moore, 
``Mechanics of Billiards, and Analysis of Willie Hoppe's Stroke," (1942).

\bibitem{Walllace88}
R. Evan Wallace and Michael C. Schroeder, 
``Analysis of billiard ball collisions in two dimensions,"
Am. J. Phys. {\bf 56}, 815 (1988); doi:10.1119/1.15455 .

\bibitem{Crown}
S. C. Crown,
``Modeling the Effects of Velocity, Spin, Frictional Coefficient, and Impact Angle on Deflection Angle in Near-elastic Collisions of Phenolic Resin Spheres,''


\bibitem{Han2005}
Inhwan Han, 
``Dynamics in Carom and Three Cushion Billiards,"
Journal of Mechanical Science and Technology {\bf 19}, 976 (2005). 

\bibitem{Mathavan1}
MATHAVAN, S., JACKSON, M.R. and PARKIN, R.M.,  
 ``Application of high-speed imaging to determine the dynamics of billiards,''
Am. J. Phys. {\bf 77} (9), 788 (2009). doi:10.1119/1.3157159.

\bibitem{Mathavan2010}
MATHAVAN, S., JACKSON, M.R. and PARKIN, R.M.,  
``A theoretical analysis of billiard ball dynamics under cushion impacts,'' 
Proceedings of the Institution of Mechanical Engineers, Part C: 
Journal of Mechanical Engineering Science, {\bf 224} (9), pp.1863 - 1873 (2010).

\bibitem{Nierhoff}
T. Nierhoff, K. Heunisch and S. Hirche, "Strategic play for a pool-playing robot," 2012 IEEE Workshop on Advanced Robotics and its Social Impacts (ARSO), Munich, Germany, 2012, pp. 72-78, doi: 10.1109/ARSO.2012.6213402;
T. Nierhoff, K. Leibrandt, T. Lorenz and S. Hirche, "Robotic Billiards: Understanding Humans in Order to Counter Them," in IEEE Transactions on Cybernetics, {\bf 46}, no. 8, pp. 1889-1899, Aug. 2016, doi: 10.1109/TCYB.2015.2457404.

\bibitem{Alciatore TP21}
D. Alciatore, TP2-1, https://billiards.colostate.edu/

\bibitem{colostate}
See, for example,  $<$billiards.colostate.edu$>$ and $<$www.sfbilliards.com/articles/BD-articles.html$>$.

\bibitem{Shepard}
Ron Shepard, "Everything You Always Wanted to Know About Cue Ball Squirt, But Were Afraid to Ask,"  $<$www.sfbilliards.com/Shepard-squirt.pdf$>$ or $<$billiards.colostate.edu$>$, (2001).


\bibitem{Cross2008}
R. Cross,  
``Cue and ball deflection (or ``squirt'') in billiards",
Am. J. Phys. {\bf 76}, 205 (2008); doi: 10.1119/1.2825392 .

\bibitem{Cross2002}
R. Cross, 
“Grip-slip behavior of a bouncing ball”, 
American Journal of Physics,  {\bf 70} (11), 1093, (2002); doi:10.1119/1.1507792.

\bibitem{Stronge2018}
Stronge, W. (2018). Impact Mechanics (2nd ed.). Cambridge: Cambridge University Press. doi:10.1017/9781139050227  .


\bibitem{hk1}
H.-C. Kim, 
``Motions of a billiard ball after a cue stroke," New Physics: Sae Mulli, 72, 208 (2022). 


\bibitem{Peskin}
C. S. Peskin, 
"Collision of Billiard balls in 3D with spin and friction", (2020).
\href{https://math.nyu.edu/~peskin/Collision_of_Billiard_Balls_in_3D_with_Spin_and_Friction.pdf}{Collision of Billiard Balls in 3D with Spin and Friction.pdf}

\end{thebibliography}
\end{document}